# Harmonics-assisted optical phase amplifier

**Wu-Zhen Li[1, 2, 3, 4], Chen Yang[1, 2, 3, 4], Zhi-Yuan Zhou[1, 2, 3*], Yan Li[1, 2, 3], Yin-Hai Li[1, 2, 3], Su-Jian Niu[1, 2, 3], Zheng Ge[1, 2, 3], Li Chen[1, 2, 3], Guang-Can Guo[1, 2, 3] and Bao-Sen Shi[1, 2, 3*]**

[1] *CAS Key Laboratory of Quantum Information, University of Science and Technology of China, Hefei, Anhui 230026, China*

[2] *CAS Center for Excellence in Quantum Information and Quantum Physics, University of Science and Technology of China, Hefei 230026, China*

[3] *Hefei National Laboratory, University of Science and Technology of China, Hefei 230088, China*

[4] *These two authors contributed equally to this article.*

[*] *e-mail: zyzhouphy@ustc.edu.cn and drshi@ustc.edu.cn*

**Abstract:** The change in the relative phase between two light fields serves as a basic principle for the measurement of the physical quantity that guides this change. It would therefore be highly advantageous if the relative phase could be amplified to enhance the measurement resolution. One well-known method for phase amplification involves the use of the multi-photon number and path entangled state known as the NOON state; however, a high-number NOON state is very difficult to prepare and is highly sensitive to optical losses. Here we propose and experimentally demonstrate in principle a phase amplifier scheme with the assistance of a harmonic generation process. The relative phase difference between two polarization modes in a polarized interferometer is amplified coherently four times with cascaded second-harmonic generation processes. We demonstrate that these amplification processes can be recycled and therefore have the potential to realize much higher numbers of multiple amplification steps. The phase amplification method presented here shows considerable advantages over the method based on NOON states and will be highly promising for use in precision optical measurements.

## Introduction

Phase is one of the most important parameters in both wave optics and quantum mechanics. The relative phase between two light fields in an interferometer or in a superposition state in quantum mechanics is highly significant because the dynamic changes in many physical quantities, including displacement, temperature, and electrical and magnetic fields, can be transduced into changes in the relative phase between light fields or wave functions[1-6]. Therefore, most high-precision measurement tasks can be converted into the measurement of the phase change in a specific physical process, and methods to amplify the phase are highly important for phase measurement resolution enhancement in metrology operations.

One well-known method that is used in quantum optics to realize phase amplification is based on the multi-photon number and path entangled state known as the NOON state $(|N::0\rangle = (|N0\rangle_{AB} + |0N\rangle_{AB})/\sqrt{2})$, which contains $N$ indistinguishable particles in an equal superposition, with all particles being in either of the paths $A$ or $B$[1]. For interference with NOON states, the phase oscillation is $N$ times faster than that for a single photon, and because the effective $N$-photon de Broglie wavelength is $\lambda / N$, this will result in phase super-resolution for high-precision measurements[7].

Although NOON states can be used for phase amplification and precision metrology applications, it is very difficult to prepare NOON states with high photon numbers; the highest number of the NOON state to be prepared to date[8] is around 10. Additionally, the detection probability is very low when $N$ is large, and a high-photon-number NOON state is highly sensitive to any optical losses experienced by the photons. Therefore, the determination of another phase amplification method that is much more robust than that using NOON states would be highly significant.

Here we propose and experimentally demonstrate, in principle, a new phase amplification scheme based on assistance from harmonic generation processes. In the proposed scheme, the relative phase between two polarization modes in a polarized interferometer is amplified coherently four times based on the cascaded second-harmonic generation (SHG) processes. Furthermore, we experimentally demonstrate that the phase amplification is not determined by the change in the laser frequency, it can still work even after the second-harmonic frequency is converted back to the fundamental frequency via the difference frequency generation (DFG) process, therefore, higher amplification levels could be achieved in an SHG-DFG recycled scheme.

Our proposed phase amplifier is realized with the assistance of a three-wave mixing process (TWM), e.g., SHG, sum-frequency generation (SFG) or DFG. In TWM, the annihilation of two photons serves to create a new photon from a microscopic viewpoint. The interaction Hamiltonian can be expressed as[9]

$$\hat{H}_{\text{eff}} = i\hbar\kappa\left(\hat{a}_1\hat{a}_2\hat{a}_3^{+} - \hat{a}_1^{+}\hat{a}_2^{+}\hat{a}_3\right) \tag{1}$$

where $\hat{a}_i$ and $\hat{a}_i^{+}$ $(i = 1,2,3)$ represent the annihilation and creation operators, respectively, of the three interacting photons; in addition, $\kappa$ is a constant that is proportional to the second-order susceptibility $\chi^{(2)}$, the pump power and other experimental parameters of the nonlinear crystal used here. In an SHG process, the annihilation of two photons at a fundamental wavelength creates a photon at the second harmonic wavelength; this simple interaction process will lead to new elements being used to manipulate different aspects of light.

This process is inspired by our previous work, in which we realized SHG of light carrying superposed orbital angular momentum modes[10]; in that work, we demonstrated that the orbital angular momentum carried by light is doubled after the SHG process and also found that the relative phase between the superposed orbital angular momentum modes is also doubled in SHG (see equation (3) in Ref. 10). Two basic questions then naturally occur: can this phase doubling effect be generalized to optical superposed modes in other degrees of freedom, such as the polarization or the optical path? Can the higher phase amplification be realized by using cascaded TWM processes? Our answer to these questions is “yes” in both cases, and in the text below, we will illustrate the basic idea of the scheme and present a detailed proof-of-principle experimental demonstration.

Figure 1 shows a graphical summary of the main concept of this work. Figure 1(a) shows a schematic of a linear Mach-Zehnder interferometer (MZI) and figure 1(b) shows a nonlinear MZI based on the use of SHG. If we assume that the same phase difference is introduced into both interferometers at the fundamental wavelength, then the speed of variation of the interference fringes in the nonlinear MZI is twice as fast as that in the linear MZI. In wave optics, the interference fields at the output ports of these linear and nonlinear MZIs can be expressed as:

$$E_{\text{L}}(\omega) = E_1(\omega) + e^{i\Phi(\omega)}E_2(\omega) \tag{2}$$

$$E_{\text{NL}}(2\omega) = E_1(2\omega) + e^{i2\Phi(\omega)+i\Delta\Phi(2\omega)}E_2(2\omega) \tag{3}$$

Where $E_1(\omega)$ and $E_2(\omega)$ are the two superposed light fields at the beam splitter, and $\Phi(\omega)$ is the phase difference between these two fields at the fundamental wavelength $E_1(2\omega)$ and $E_2(2\omega)$ are the two superposed SHG light fields at the beam splitter, and the relationships between the SHG beam and the two fundamental beams are given by $E_1(2\omega) \propto E_1(\omega)^2$ and $E_2(2\omega) \propto E_2(\omega)^2$, respectively; and $\Delta\Phi(2\omega)$ is the phase difference between the SHG beams. If the nonlinear interferometer contains $n^{\mathrm{th}}$ harmonic generation processes, then the output fields from the nonlinear MZI can be expressed as:

$$E_{\mathrm{NL}}(n\omega) = E_1(n\omega) + e^{in\Phi(\omega)+i\Delta\Phi(n\omega)}E_2(n\omega) \quad (4)$$

where $E_1(n\omega)$ and $E_2(n\omega)$ are the two superposed $n^{\mathrm{th}}$ harmonic light fields at the beam splitter, and the relationships between these $n^{\mathrm{th}}$ harmonic beams and the fundamental beam are given by $E_1(n\omega) \propto E_1(\omega)^n$ and $E_2(n\omega) \propto E_2(\omega)^n$, respectively. $\Delta\Phi(n\omega)$ represents the total phase difference that contains the phase difference of the harmonic beams ranging from the 2nd to the $n$th. These phase differences are caused by different dispersion of the two polarized modes transmitting through different optical elements such as nonlinear crystal, wave plates, polarizing beam splitters (PBSs), etc., and can be kept constant during the experiments. By assuming $E_1(\omega) = E_2(\omega)$, equations (3) and (4) clearly indicate that the phase difference $\Phi(\omega)$ introduced by the fundamental beam is amplified to $n\Phi(\omega)$ and the light intensity is given by $I_1(n\omega)\{1+\cos[n\Phi(\omega)+\Delta\Phi(n\omega)]\}$. The dependence on $n\Phi(\omega)$ rather than $\Phi(\omega)$ is phase super-resolution: one cycle of $I_{\mathrm{NL}}(n\omega)$ implies a smaller change of the phase difference than one cycle of $I_{\mathrm{L}}(\omega)$ [1].

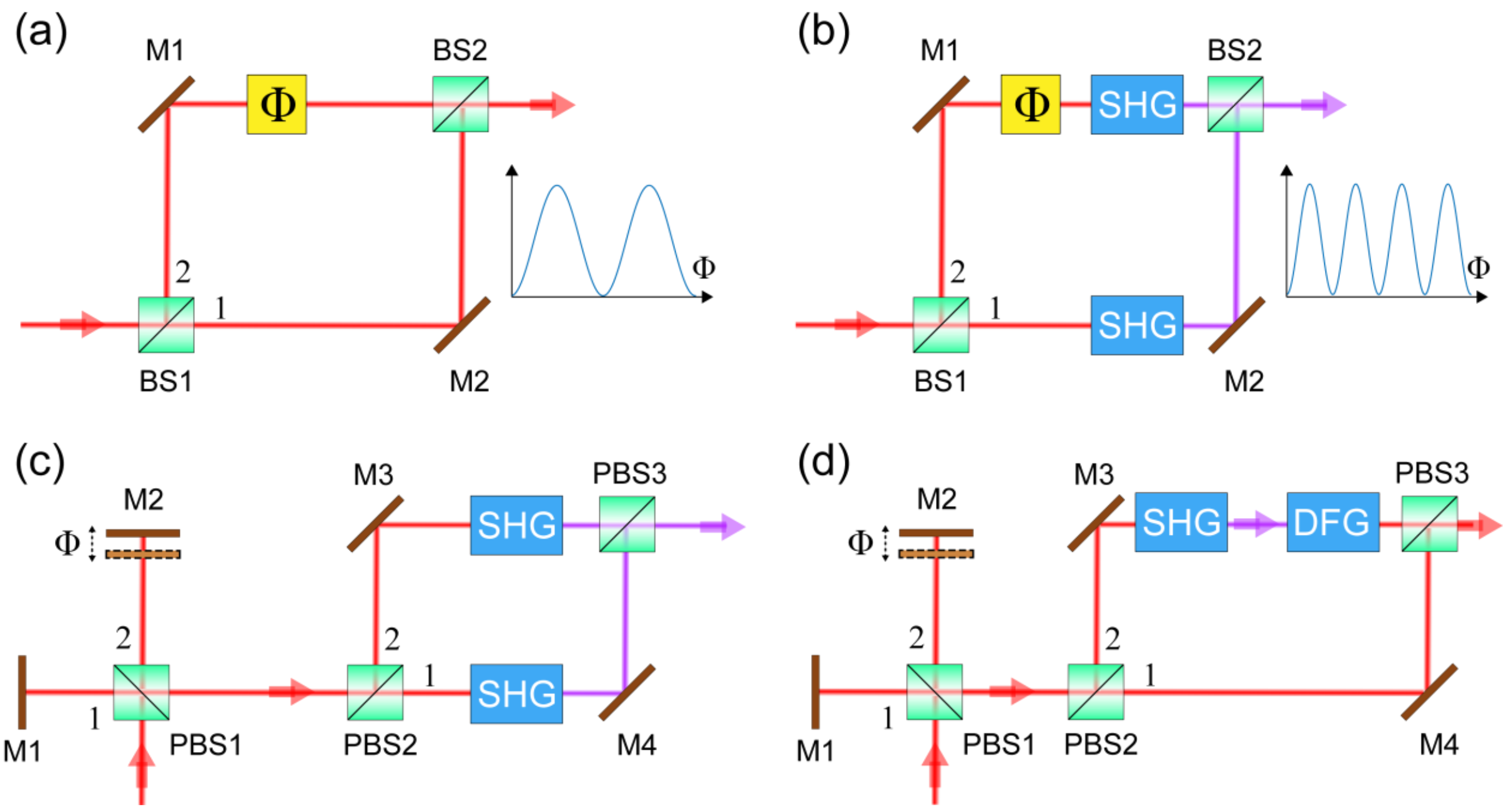

Fig. 1. Schematics of the interferometer configurations. (a) MZI. (b) Nonlinear MZI. (c) Nonlinear polarized interferometer. (d) Configuration for phase doubling without frequency doubling.

In figure 1(c), a polarized Michelson interferometer is used instead of the MZI and works in a manner similar to the system in figure 1(b), but it is much easier to realize experimentally. For the scheme in figure 1(c), $\Phi(\omega)$ in Eq. (3) can be adjusted by controlling arm difference in the Michelson interferometer with a modulator or a piezoelectric transducer, while $\Delta\Phi(2\omega)$ is a constant that contains all the phase difference generated in the MZI.

To demonstrate that the phase amplification is not frequency-dependent, we propose and experimentally demonstrate the scheme shown in figure 1(d). In this scheme, a SHG-DFG module that can double the phase but not change the frequency is inserted in one arm of the MZI, which converts the frequency from $\omega$ to $2\omega$ in the SHG process and from $2\omega$ back to $\omega$ in the DFG process. The interference field at the output ports can be expressed as

$$E_{\mathrm{NL}}(\omega) = E_1(\omega) + e^{\mathrm{i}2\Phi(\omega)+\mathrm{i}\Delta\Phi(2\omega)} E_2(\omega) \quad (5)$$

The phase difference is amplified by 2 times without changing the frequency, which demonstrates that the phase amplification is frequency-independent. Moreover, if the nonlinear conversion efficiencies for both SHG and DFG are high enough, then the output light can enter another SHG-DFG module to realize 4 times phase amplification. Considering that the nonlinear conversion efficiency is dependent on the laser intensity, this scheme could be cascaded more times to achieve much higher amplification levels for the phase difference if a high-intensity laser is used. Some additional recycling methods that can be used to achieve high amplification levels are described in the Supplementary Information.

## Results

To realize interference of the SHG beams, a phase doubling module can be constructed using either a Sagnac loop with a nonlinear crystal[11] or orthogonally placed nonlinear crystals[12]. These two methods represent better options than the MZI with respect to the phase difference stability because the two polarization components transmit through the same path. In our experiment, the Sagnac loops and orthogonally placed crystals are cascaded to realize interference of two fourth-harmonic (FH) beams and thus demonstrate that the measurement resolution of the phase difference is amplified by four times.

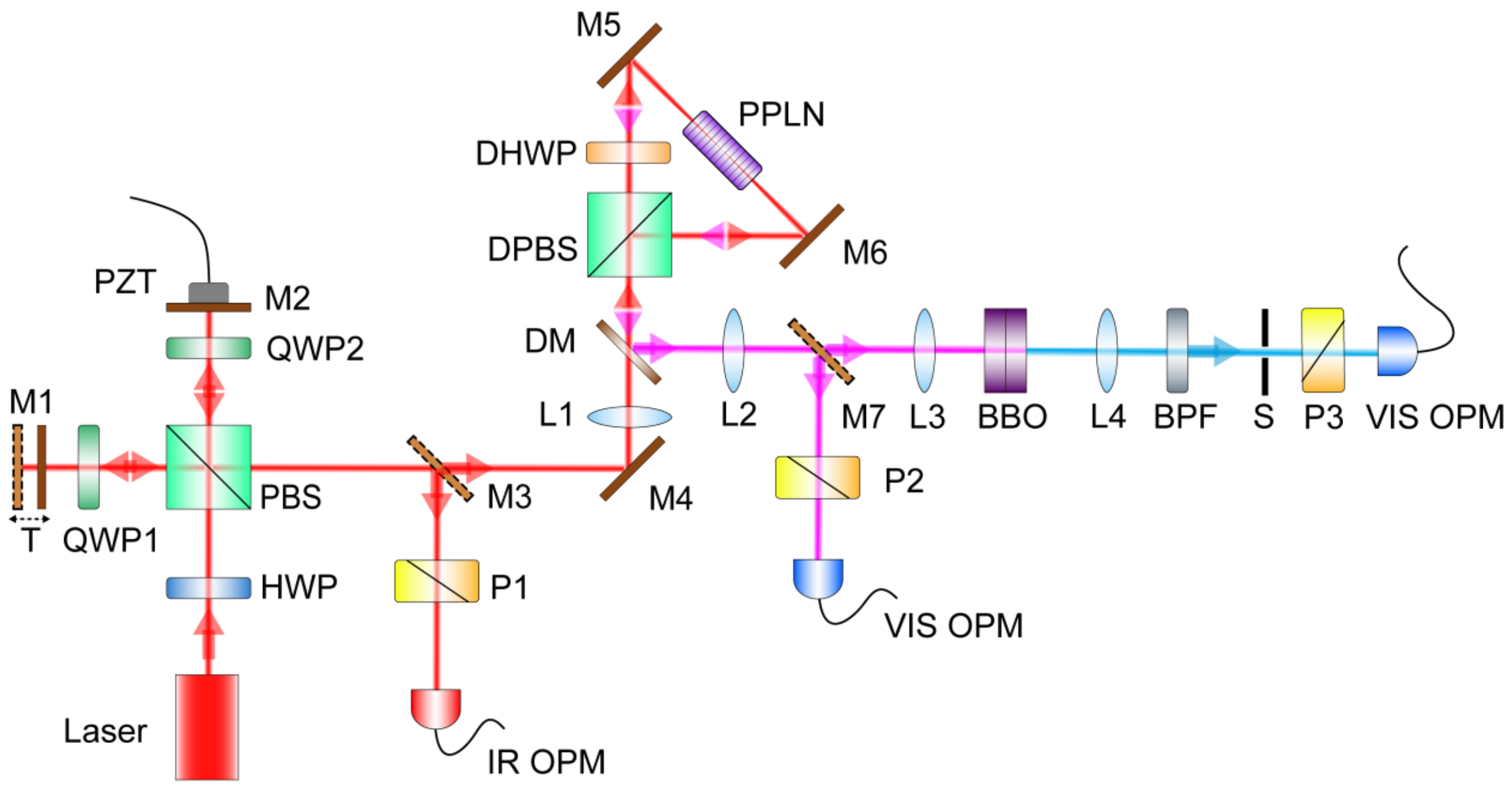

Fig. 2. Experimental setup schematic. HWP: half-wave plate; QWP: quarter-wave plate; DHWP: 1560 nm/780 nm dichroic HWP; PBS: polarizing beam splitter; DPBS: dichroic PBS; M: silver-coated mirror; T: translation stage; PZT: piezoelectric transducer; DM: dichroic mirror; BPF: 390–10 nm band-pass filter; PPLN: periodically poled lithium niobate crystal; BBO: β-barium borate crystal; S: small hole; P: polarizer; IR OPM: infrared optical power meter; VIS OPM: visible optical power meter; L: lens; lenses L1 and L3 are used for focusing and L2 and L4 are used for collimation, and the focal lengths of L1–L4 are 200 mm, 200 mm, 50 mm, and 100 mm, respectively.

We used a femtosecond pulsed laser with a centre wavelength of 1560 nm as the fundamental frequency light source to pump the two cascaded SHG modules, as shown in figure 2. Full details of the experimental setup are summarized in the Methods section. The interference results for the fundamental, SH and FH beams are shown in figure 3(a), (b) and (c), respectively. The *y*-axis in figure 3 represents the measured optical power and the *x*-axis represents the change in the optical path difference (OPD). As shown in figure 3, the periods of the interference curves of the SH and FH beams are reduced to 1/2 and 1/4, respectively, compared to the fundamental interference curves, which indicates that the phase difference 2π corresponding to one period of the interference curve is amplified to 4π and 8π respectively. Therefore, these results clearly demonstrate that the phase difference for the fundamental beam has been doubled and quadrupled by SHG and FHG, respectively.

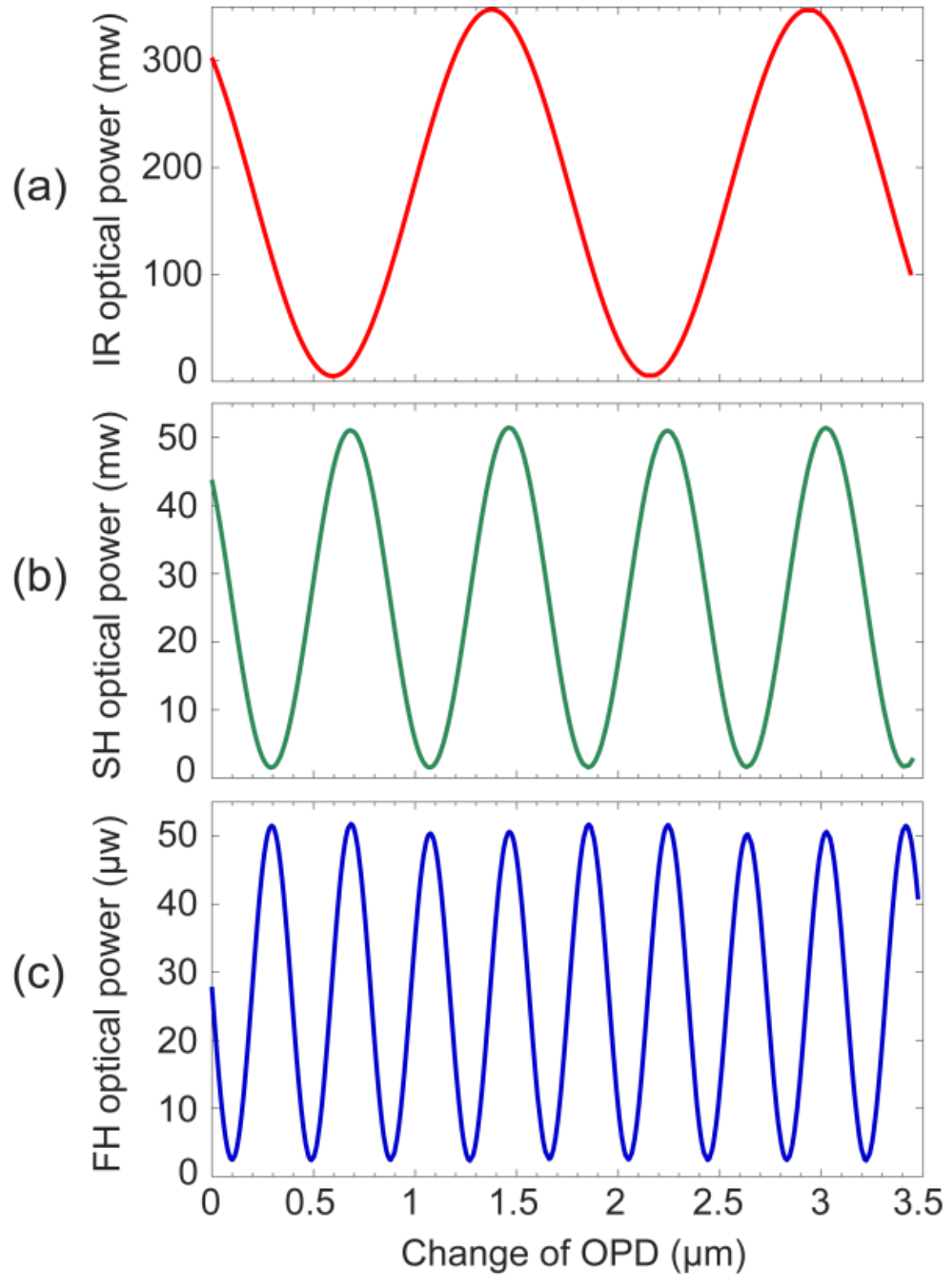

Fig. 3. Dependence of optical power after interference on the change in OPD. (a) Fundamental light; (b) SH light; (c) FH light.

Significantly, the three OPDs shown in figure 3(a)–(c) are all obtained from the same Michelson interferometer, in which the centre wavelength of the light is always 1560 nm. Although the observed phase doubling is accompanied by frequency doubling, the latter is not the reason for the occurrence of the former.

To demonstrate that the observed phase doubling is independent of the operating wavelength, we performed a DFG process to convert light with the wavelength of 780 nm back to 1560 nm and thus verify that the phase doubling still exists. Because the output light after both SHG and DFG has the same wavelength as the original input light, phase doubling can be realized by implementing both SHG and DFG in a single arm of the interferometer. In the experiment, the DFG procedure is the inverse process of type-II (yzy) SHG; the setup used is shown in figure 4(a). We assume here that the states of the horizontally and vertically polarized light after the Michelson interferometer are $|\phi_1,\omega,\mathrm{H}\rangle$ and $|\phi_2,\omega,\mathrm{V}\rangle$, respectively, where $\phi_1$ and $\phi_2$ represent the initial phases, which are dependent on the lengths of the two interferometer arms. After SHG, the state $|2\phi_2,2\omega,\mathrm{V}\rangle$ is generated from $|\phi_2,\omega,\mathrm{V}\rangle$. A dichroic mirror (DM) is used to separate the light at the different wavelengths (1560 nm and 780 nm) into different paths. In the path for the 1560 nm light, a polarizing beam splitter is used to filter out the state $|\phi_2,\omega,\mathrm{V}\rangle$. The two remaining states, $|2\phi_2,2\omega,\mathrm{V}\rangle$ and $|\phi_1,\omega,\mathrm{H}\rangle$, are then combined to perform the DFG operation and the state $|2\phi_2-\phi_1,\omega,\mathrm{V}\rangle$ is generated. Finally, a polarized Michelson interferometer is used to compensate for the OPD caused by polarization mode dispersion, and a polarizer is used to produce polarized light interference between the states $|2\phi_2-\phi_1,\omega,\mathrm{V}\rangle$ and $|\phi_1,\omega,\mathrm{H}\rangle$. The interference result is dependent on the phase difference $2(\phi_2-\phi_1)$, which is doubled compared with the original phase difference. The data shown in figure 4(b) confirm that the phase difference is doubled.

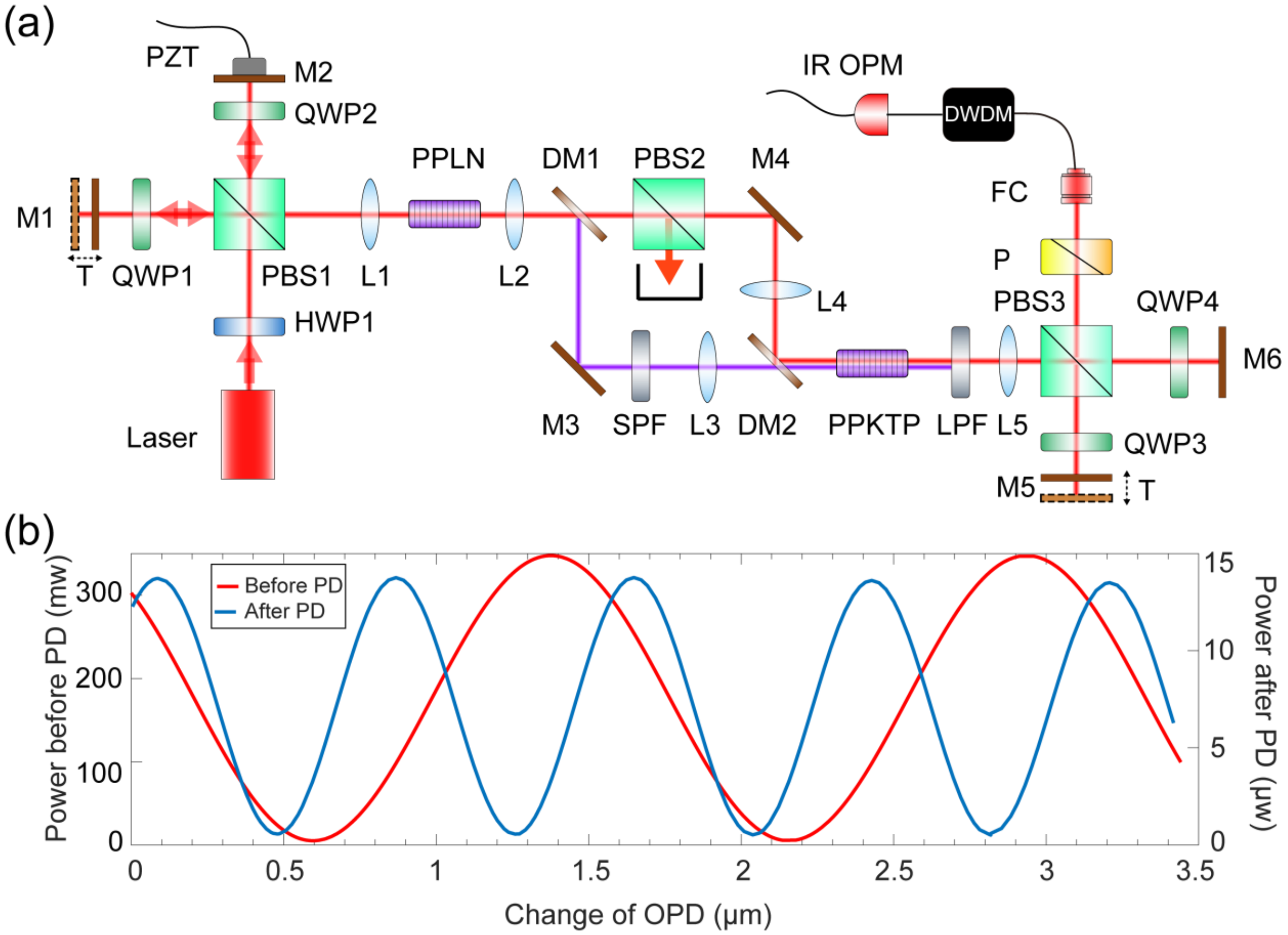

Fig. 4. Experimental setup and results. (a) Experimental setup used for phase doubling (PD) without frequency doubling. SPF: 1000 nm short pass filter; LPF: 1400 nm long pass filter; FC: fibre collimator; CH22: channel 22 of a dense wavelength division multiplexing system with a centre wavelength of 1559.79 nm. (b) Measured optical power characteristics of interference before and after phase doubling.

## Discussion

In summary, we have presented an attractive method to enable the realization of an optical phase amplifier with the assistance of a TWM process in an interferometer. We achieve a phase difference amplification of four times in the optical phase amplifier by demonstrating the reduction of the period of the interference curve to 1/4 of the original period. This indicates that anything causing the interference curve to change for one period can cause the interference curve to change for 4 periods by using the optical phase amplifier, therefore the resolution of the interferometer is improved by 4 times. We believe the optical phase amplifier presented in this work could be used in numerous precision measurement scenarios, such as the measurements of the optical properties (for example, dispersion and absorption coefficients) of transparent materials[13,14], other physical quantities such as displacements, angles, and electrical and magnetic fields, which can be transduced into changes in an optical path length[15], or interference-based imaging for enhanced image resolution[16], etc. In principle, the proposed optical phase amplifier that is based on SHG in our experiment could be generalized to high-order harmonic generation processes[17,18].

Although a maximum phase amplification of four is achieved in the experiments presented here, a SHG-DFG recycling scheme described in detail in the supplementary material that provides the potential to achieve higher phase amplification levels is proposed too. In our experiment, the FHG efficiency of 0.07% and the DFG efficiency of ~ 2% are low because of the low pump power and mismatched bandwidths (the bandwidths of two input lights and the acceptance bandwidth of the crystal should be matched to obtain high DFG efficiency). Fortunately, under suitable experimental conditions (with a high-intensity pump laser and a proper crystal), the power conversion efficiency of both SHG and DFG can be improved up to 80%[19,20], therefore, much higher phase amplification levels could be achieved in principle. If the efficiencies of the nonlinear processes are high enough, the amplification level is expected to surpass the highest amplification levels achieved using NOON states. Moreover, the method presented in this work is much more robust than that based on NOON states because the latter needs to take a long time in the coincidence measurement by using single-photon detectors. While in this work, the high-speed photodiode can be used in the real-time measurement, the interference curves are close to perfect sine curves, error bars and fitting are not necessary which are important for processing interference results of NOON states[1,21,22].

We realize that there are various studies in phase estimation[23-26]. We would like to clarify the obvious differences between the present work and those schemes. Previous literatures focus on the phase estimation of non-classical light sources in a nonlinear interferometer. The light source generated is far below the threshold of nonlinear processes and the mean photon number is very small, so these works study the phase estimation in the quantum regime and one can achieve Heisenberg limits (1/N) for phase estimation. Besides, the phase estimation is based on the coincidence measurement by using single-photon detection or homodyne detection. While in our present scheme, all light sources are lasers, which are above the threshold of the nonlinear processes, the mean photon number is very large, a high-speed photodiode can be used for measuring the interference curve, and the limits for phase measurement are the standard quantum limit ($1/\sqrt{N}$). The distinct feature of our present scheme is the phase super-resolution: after our phase amplifier, a change in light intensity in one cycle corresponds to a much smaller change in phase difference than that before the phase amplification. The work presented here will revolutionize our understanding of nonlinear interference and may open new avenues for interference-based precision metrology.

## Methods

**FHG experiment.** A schematic of the experimental setup is shown in figure 2. The input light is a pulsed laser beam with a centre wavelength of 1560 nm, a repetition frequency of 80 MHz and a pulse width of 150 fs. The linearly polarized beam is transformed into a 45°-polarized beam using a half-wave plate (HWP), and thus the input beam is divided equally into two orthogonal linearly polarized light beams by the PBS. In each arm of the Michelson interferometer, a group composed of a mirror and a quarter-wave plate (QWP) can transform the beam's polarization state into its orthogonal state[10], which causes the light to exit the interferometer from the other port of the PBS. Mirror M1 is fixed on a translation stage that is used to perform rough adjustments of the OPD between the two arms. Mirror M2 is fixed on a piezoelectric transducer (PZT) that is used to enable fine adjustment of the OPD within a 3.5 μm range; the PZT is driven using an amplified triangular wave signal.

In the first SHG module, a periodically poled lithium niobate (PPLN) crystal with a length of 5 mm is used. The PPLN crystal satisfying the type-0 (zzz) quasi-phase-matching condition only responds to the vertical polarization of light. Here, a Sagnac loop with a dichroic HWP inserted is used to realize the SHG process for both vertical and horizontal polarizations, which was demonstrated in our previous work[11].

In our experiment, the average pump power in each arm is 135 mW which is limited by the optimum output power of the femtosecond laser; SH light with a power of 30 mW at a centre wavelength of 780 nm is generated in each polarization state. The second SHG module is composed of a crystal group that comprises two orthogonally glued type-I (ooe) birefringence phase matching (BPM)-type β-barium borate (BBO) crystals. Each BBO crystal has a thickness of 0.5 mm and a phase-matching angle of ~ 30°. The two 780 nm orthogonally polarized light beams are frequency-doubled by the two orthogonally placed BBO crystals. Each polarization state of the 390 nm SH light has an optical power of 95 μW. Therefore, the two orthogonally polarized 1560 nm beams can be frequency-doubled twice sequentially by cascading two SHGs, and the resulting 390 nm light can be considered to be the FHG output from the input 1560 nm laser beam.

The interference of the FH orthogonal polarization states can be observed after the light passes through a 45° polarizer. An optical power meter is used to record the change in the power that occurs when the OPD in the Michelson interferometer is scanned using the PZT driven by the amplified triangular wave. When we observed and recorded the interference of the fundamental frequency light or the SH light, mirror M3 or M7 was inserted, respectively, to separate the light beam from its original optical path; otherwise, these two mirrors are not present in the optical path.

**DFG experiment.** In the experiments performed to demonstrate that the phase amplification process is independent of the operating wavelength, we introduced the DFG process, as illustrated in figure 4(a). In this experiment, we used the same Michelson interferometer configuration and the same PPLN crystal for frequency doubling used in the FHG experiment described earlier. Here, the Sagnac loop is not used, so the crystal only responds to the vertical polarization of light. DM1 transmitted the 1560 nm light and reflected the 780 nm light, while DM2 had the opposite characteristics. The crystal that was used for DFG was a type-II (yzy) periodically poled potassium titanyl phosphate (PPKTP) crystal with a length of 10 mm. A PBS and a short pass filter (SPF) were used to filter out the 1560 nm light, thus suppressing the background noise from the difference frequency light and improving the visibility of the interference. The second Michelson interferometer was used to compensate for the OPD caused by polarization mode dispersion. Because the spectrum of the 1560 nm light after SHG and DFG was changed when compared with the original spectrum, a dense wavelength division multiplexing system was used as a narrow band filter (full width at half maximum of 200 GHz) to improve the interference visibility.

## Acknowledgements

This work is supported by the National Natural Science Foundation of China (NSFC) (11934013, 92065101), Anhui Initiative in Quantum Information Technologies (AHY020200), and Innovation Program for Quantum Science and Technology (2021ZD0301100).

## Conflict of interest

The authors declare no competing financial interests.

## Author contributions

Z.-Y.Z. and C.Y. designed the experiment. W.-Z.L. and C.Y. carried out the experiment with assistance from Y.L. and Y.-H.L. S.-J.N., Z.G. and L.C. helped collect the data. C.Y., Z.-Y.Z. and W.-Z.L. analyzed the data and wrote the paper with input from all other authors. The project was supervised by Z.-Y.Z., B.-S.S. and G.-C.G. All authors discussed the experimental procedures and results.

## Supplementary information

The online version contains supplementary material available at https://doi.org/10.1038/s41377-022-01003-3.

## References

1 Giovannetti, V., Lloyd, S. & Maccone, L. Quantum-enhanced measurements: beating the standard quantum limit. *Science* **306**, 1330-1336 (2004).

2 Thompson, A. R., Moran, J. M. & Swenson, G. W. Jr. Interferometry and Synthesis in Radio Astronomy. 3rd edn. (Cham: Springer, 2017).

3 Saulson, P. R. Fundamentals of Interferometric Gravitational Wave Detectors. (River Edge: World Scientific, 1994).

4 Williams, D. C. Optical Methods in Engineering Metrology. (Dordrecht: Springer Science & Business Media, 2012).

5 Nolte, D. D. Optical Interferometry for Biology and Medicine. (New York: Springer Science & Business Media, 2012).

6 Brezinski, M. E. Optical Coherence Tomography: Principles and Applications. (Amsterdam: Academic Press, 2006).

7 Jacobson, J. *et al*. Photonic de broglie waves. *Physical Review Letters* **74**, 4835-4838 (1995).

8 Wang, X. L. *et a*l. Experimental ten-photon entanglement. *Physical Review Letters* **117**, 210502 (2016).

9 Kumar, P. Quantum frequency conversion. *Optics Letters* **15**, 1476-1478 (1990).

10 Zhou, Z. Y. *et al*. Orbital angular momentum light frequency conversion and interference with quasi-phase matching crystals. *Optics Express* **22**, 20298-20310 (2014).

11 Yang, C. *et al*. Nonlinear frequency conversion and manipulation of vector beams in a Sagnac loop. *Optics Letters* **44**, 219-222 (2019).

12 Liu, H. G. *et al*. Nonlinear frequency conversion and manipulation of vector beams. *Optics Letters* **43**, 5981-5984 (2018).

13 Tian, C. S. & Shen, Y. R. Recent progress on sum-frequency spectroscopy. *Surface Science Reports* **69**, 105-131 (2014).

14 Chekhova, M. V. & Ou, Z. Y. Nonlinear interferometers in quantum optics. *Advances in Optics and Photonics* **8**, 104-155 (2016).

15 Abbott, B. P. *et al*. Observation of gravitational waves from a binary black hole merger. *Physical Review Letters* **116**, 061102 (2016).

16 Meek, A. T. *et al*. Real-time imaging of cellular forces using optical interference. *Nature Communications* **12**, 3552 (2021).

17 Lv, Y. Y. *et al*. High-harmonic generation in Weyl semimetal β-$WP_2$ crystals. *Nature Communications* **12**, 6437 (2021).

18 Ghimire, S. *et al*. Observation of high-order harmonic generation in a bulk crystal. *Nature Physics* **7**, 138-141 (2011).

19 Aparajit, C. *et al*. Efficient second-harmonic generation of a high-energy, femtosecond laser pulse in a lithium triborate crystal. *Optics Letters* **46**, 3540-3543 (2021).

20 Murray, R. T. *et al*. Highly efficient mid-infrared difference-frequency generation using synchronously pulsed fiber lasers. *Optics Letters* **41**, 2446-2449 (2016).

21 Mitchell, M. W., Lundeen, J. S. & Steinberg, A. M. Super-resolving phase measurements with a multiphoton entangled state. *Nature* **429**, 161-164 (2004).

22 Walther, P. *et al*. De Broglie wavelength of a non-local four-photon state. *Nature* **429**, 158-161 (2004).

23 Luis, A. Nonlinear transformations and the Heisenberg limit. *Physics Letters A* **329**, 8-13 (2004).

24 Pezzé, L. & Smerzi, A. Entanglement, nonlinear dynamics, and the heisenberg limit. *Physical Review Letters* **102**, 100401 (2009).

25 Cheng, J. Quantum metrology for simultaneously estimating the linear and nonlinear phase shifts. *Physical Review A* **90**, 063838 (2014).

26 Jiao, G. F. *et al*. Nonlinear phase estimation enhanced by an actively correlated Mach-Zehnder interferometer. *Physical Review A* **102**, 033520 (2020).

# Supplementary Information for

# Harmonics-assisted optical phase amplifier

Wu-Zhen Li[1, 2, 3, 4], Chen Yang[1, 2, 3, 4], Zhi-Yuan Zhou[1, 2, 3, 5], Yan Li[1, 2, 3], Yin-Hai Li[1, 2, 3], Su-Jian Niu[1, 2, 3], Zheng Ge[1, 2, 3], Li Chen[1, 2, 3], Guang-Can Guo[1, 2, 3] and Bao-Sen Shi[1, 2, 3, 6]

[1] CAS Key Laboratory of Quantum Information, University of Science and Technology of China, Hefei, Anhui 230026, China

[2] CAS Center for Excellence in Quantum Information and Quantum Physics, University of Science and Technology of China, Hefei 230026, China

[3] Hefei National Laboratory, University of Science and Technology of China, Hefei 230088, China

[4] These two authors contributed equally to this article.

[5] zyzhouphy@ustc.edu.cn

[6] drshi@ustc.edu.cn

## 1. Data acquisition and processing

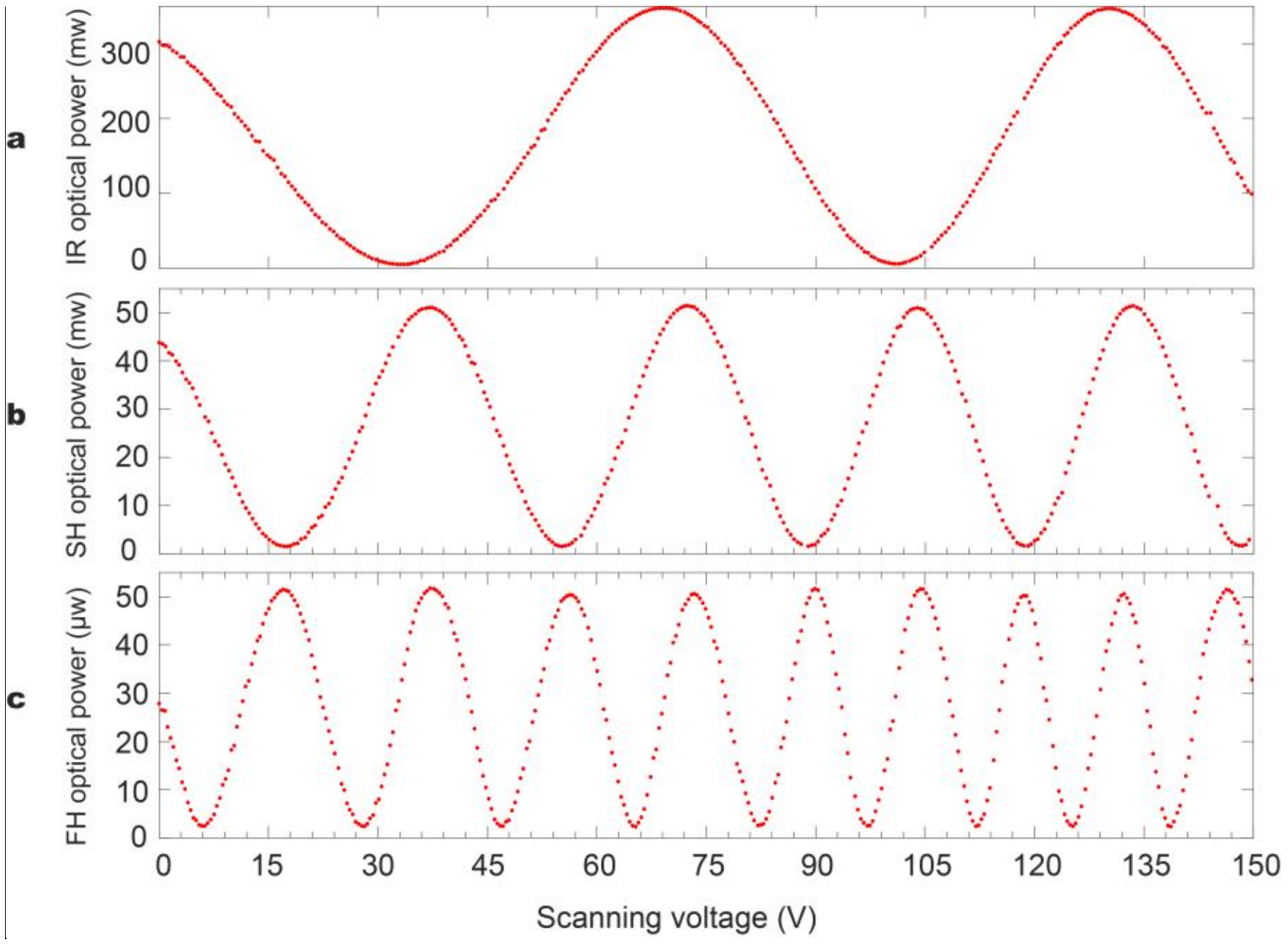

Fig. S1 Raw data with maximum interference visibility.

In the experiment, we studied the phase amplification effect of the second and fourth harmonic generation in a Michelson interferometer. Fig. S1 shows the raw interference data recorded by an optical power meter (OPM) at the three positions of the interferometer. We first use the mirror M3 in Fig. 2 (in the main text) to separate the fundamental frequency beams, and then we record the polarized light interference results using an IR OPM. The measured optical power of the fundamental frequency beams against the PZT scanning voltage is shown in Fig. S1a. In the scanning voltage range of 0 to 150 V, the optical power changes by ~2.25 periods, which indicates that the optical phase difference between the two arms of the interferometer transforms by ~ 4.5π. After recording the results based on fundamental pulse light, we removed M3 to perform the SHG experiment. Then we use M7 to separate the SHG beams, and the optical power after interference is given in Fig. S1b. For the SHG beams, the interference intensity changes ~ 4.5 periods; therefore, phase difference changes by 9π within the scanning voltage range of 150 V. We then remove M7 to perform FHG experiments based on the SHG beams, and the optical power after interference between the two FHG beams is also recorded by the VIS OPM; the raw interference data for FHG beams are shown in Fig. S1c.

Since the relationship between the scanning voltage and the displacement of the PZT is nonlinear, the arm difference ΔL does not change linearly with the voltage of the PZT. Thus, from Fig. S1, we find that the changing period of the optical power of interference with respect to the scanning voltage is a variable. In order to unify the changing period of the optical power, we need to convert the scanning voltages into changes of the optical path difference (OPD), as shown in Fig. 3 in the main text. We first perform function fitting on the experimental data in Fig. S1 to obtain an accurate function relationship between changes of OPD and voltages, and then we make use of the function to convert the scanning voltages into the corresponding changes of OPD.

In our experiment, we collected ~ 320 data points for each interference result within the 150 V scanning range. The curves shown in Fig. 3 in the main text are the results of the direct connection of the data points.

## 2. Scheme of achieving high amplification times

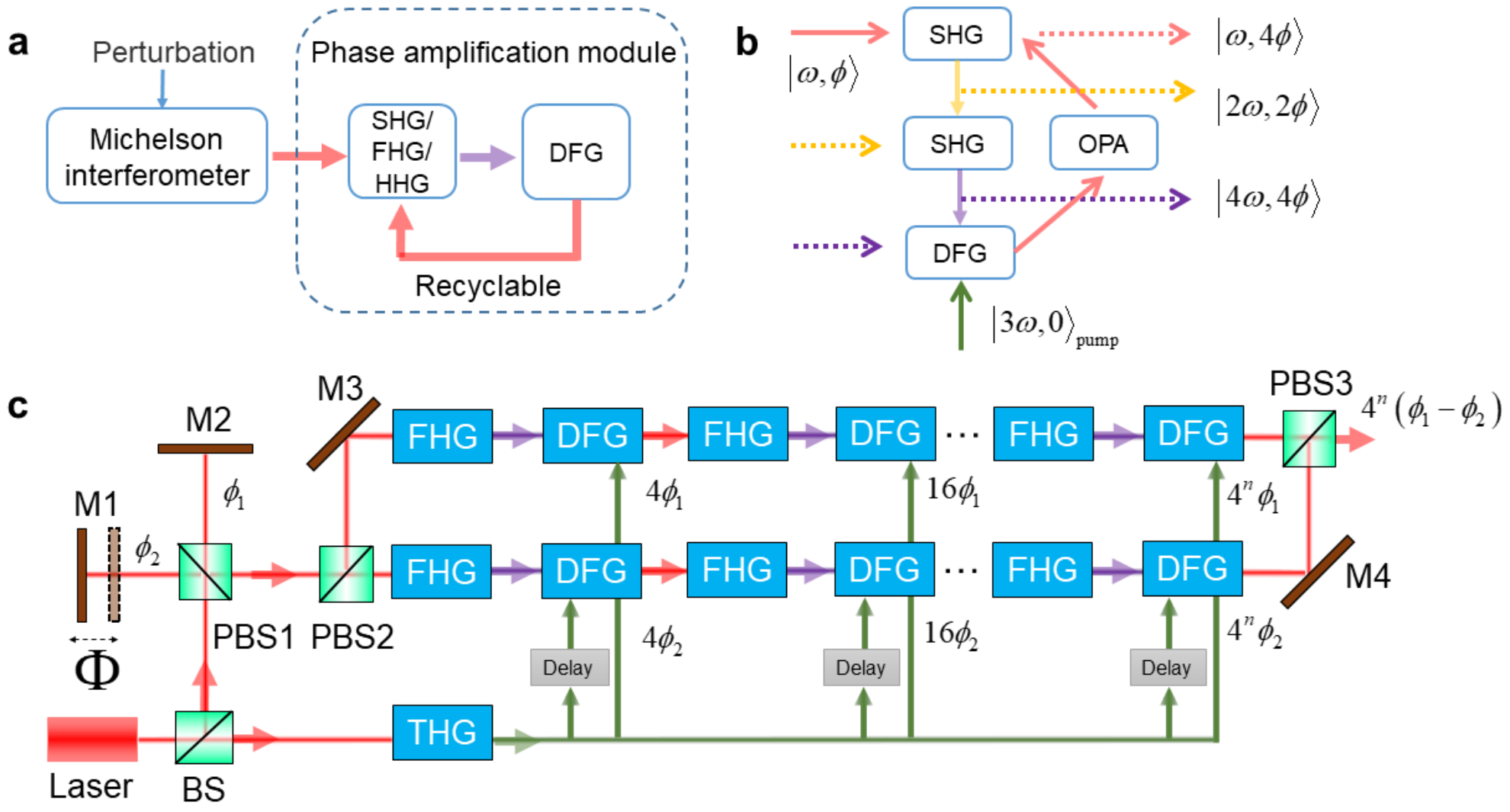

Fig. S2. Schematic of the recyclable phase amplification module. **a** A basic flow chart. **b** A schematic of the FHG-DFG recycling module. **c** An experimental scheme to realize FHG-DFG recycling. The fundamental, FH, and pump lights are marked red, purple, and green, respectively.

In the main text, we present a scheme of a phase amplifier that can amplify the phase difference in an interferometer by 4 times. We also mention that higher amplification times can be obtained by using a recycling scheme introduced in the following.

A basic flow chart is shown in Fig. S2a. Any changes in physical quantities affecting the optical path length can be sensed and transformed into changes in phase difference by a Michelson interferometer. The phase difference can be amplified by the optical phase amplifier that is formed by harmonic generation modules and difference frequency generation (DFG) modules. The amplification is the *n*-th power of the amplification in one cycle, where *n* represents the number of cycles. This flow chart includes the situation presented in the main text, where the SHG module and DFG module are cascaded. To improve the order of harmonic waves, one can cascade more SHG modules except by using high harmonic generation materials. For second-order nonlinear crystals, the complex high-order nonlinear processes are relatively weak; therefore, the noise is easy to filter.

In Fig. S2b, we present a FHG-DFG scheme where two SHG modules are cascaded and used as a fourth-harmonic generation (FHG) module. The advantage of FHG is that the high-efficiency DFG of type-0 quasi-phase-matching can be utilized. In the SHG-DFG scheme that is demonstrated in the main text, the DFG should be based on the type-II phase-matching condition, where the efficiency is much lower than that of type-0 phase-matching. In the recycling module shown in Fig. S2b, the input or output light can not only work at the fundamental frequency $\omega$, but also at SH frequency $2\omega$ or FH frequency $4\omega$, which can be flexibly designed according to the working waveband of the interferometer and the detector. A simple method to realize the recycling is to cascade the combination of FHG and DFG modules, as shown in Fig. S2c.

Unlike the scheme in the main text, an extra pump beam should be introduced in the DFG (and OPA if it is used) modules, and the extra pump beam can introduce a random phase on the DFG light. Therefore, the recycling must be implemented on both two paths and the pump beams should come from the same laser to ensure that the random phase is introduced on both paths and no extra random phase difference is generated. An easy pump scheme is proposed as follows. One can split a part of the fundamental light to implement the third-harmonic generation (THG) process (the THG can be implemented by combining an SHG process and a sum-frequency generation process) and use the THG light with a frequency of $3\omega$ as the pump light of the DFG modules. In a DFG module, the input lights are an FHG signal light ( $4\omega$ ) and a THG pump light ( $3\omega$ ), and the output light has the fundamental frequency $\omega$ . Therefore, the output can enter the next FHG-DFG module to realize higher amplification. Significantly, the phase delay between the THG pump light in two paths should be carefully adjusted to make the phase difference generated in the two DFG processes zero.

To increase the number of cycles, the loss in the loop should be low and the nonlinear efficiency should be high. In our experiment, the pulsed pump beam has a full width at half maximum (FWHM) of 150 fs and energy below 5 nJ. Since the frequency conversion efficiency of the ultrashort pulse is limited by low pump energy, bandwidth mismatch (the bandwidths of the input lights and the acceptance bandwidth of the crystal should be matched to obtain high efficiency), and group velocity mismatch in the crystal, the FHG efficiency obtained in our experiment is less than 0.1%. However, high FHG efficiency can be easily achieved by choosing appropriate pump light sources and nonlinear crystals.

For example, in Ref. 1, the FHG energy efficiency of 46.5% was obtained at the fundamental intensity of 1.5 GW $cm^{-2}$, where a large aperture Nd: glass laser that emits high energy pulses is used as the pump light source and an integrated conversion cell consisting mainly of two KDP crystals is used as the nonlinear material. Besides, the synchronously pumped scheme and the optical parameter amplification (OPA) technique can be used to enhance the optical power of the fundamental light generated from the DFG process. For example, in Ref. 2, by synchronizing the signal and pump pulses, the picosecond mid-infrared idler radiation with a high average power (>3.6 W) and high conversion efficiency (maximum pump-to-DFG power conversion 78%) was obtained from the single-pass DFG system. In 2015, Lin Xu et al. reported a high-energy picosecond OPA with an overall conversion efficiency of 45%, where the gain of signal light can reach 79 dB for low seed powers [3]. Therefore, if a high-power sub-nanometer laser is used as the pump source, which is easily obtained by utilizing chirped pulse amplification (CPA) technology, suitable nonlinear crystals are used for FHG and DFG, and OPA technique is optionally used to amplify the lights after the DFG processes, then, a high-energy pulse can be obtained after one cycle and has sufficient energy to realize FHG-DFG processes in the next cycle.

A phase amplifier based on FHG can achieve 4 times amplification of phase difference. Therefore, the two cycles scheme of FHG-DFG-FHG can realize 16 times amplification, which can easily surpass the highest amplification levels achieved using NOON states (the highest number of the NOON state to be prepared to date is 10, which is reported in Ref. 4). The cycling scheme proposed here has the potential to realize higher phase amplification, which could be verified in future experiments.

**Reference**

1 Bruneau, D., Tournade, A. M. & Fabre, E. Fourth harmonic generation of a large-aperture Nd:glass laser. *Appl. Optics* **24**, 3740-3745 (1985).

2 Murray, R. T., Runcorn, T. H., Kelleher, E. J. R. & Taylor, J. R. Highly efficient mid-infrared difference-frequency generation using synchronously pulsed fiber lasers. *Opt. Lett.* **41**, 2446-2449 (2016).

3 Xu, L., Chan, H. Y., Alam, S. U., Richardson, D. J. & Shepherd, D. P. High-energy, near- and mid-IR picosecond pulses generated by a fiber-MOPA-pumped optical parametric generator and amplifier. *Opt. Express* **23**, 12613-12618 (2015).

4 Wang, X. L. *et al.* Experimental ten-photon entanglement. *Phys. Rev. Lett.* **117** (2016).